\documentclass{article}

\usepackage{amsmath,amsfonts}

\usepackage{color}
\usepackage{algorithm, algpseudocode, algorithmicx}
\usepackage{setspace}
\usepackage{verbatim}

\usepackage[dvips]{graphicx}
\graphicspath{{./eps/}}
\DeclareGraphicsExtensions{.eps}
\usepackage{subfig}

\title{Partitioning Uncertain Workflows}

\author{Bernardo A. Huberman and Freddy C. Chua\\Mechanisms and Design Lab\\HPLabs}

\begin{document}

\maketitle

\begin{abstract}
	It is common practice to partition complex workflows into separate channels in order to speed up their completion times. When this is done within a distributed environment, unavoidable fluctuations make individual realizations depart from the expected average gains. We present a method for breaking any complex workflow into several workloads in such a way that once their outputs are joined, their full completion takes less time and exhibit smaller variance than when running in only one channel. We demonstrate the effectiveness of this method in two different scenarios; the optimization of a convex function and the transmission of a large computer file over the Internet.

\end{abstract}
\newpage
It is well known that the partition of large workflows into smaller workloads can often accelerate the completion of the full process. This is the assumption underlying the parallelization of large computer jobs, such as the use of map-reduce for indexing documents \cite{Dean2008}, parallel algorithms for machine learning \cite{Chu2007, Low2012, Xing2015, Zinkevich2010}, decentralization of load balancing in networks \cite{Brunner2009, John2013, Prieto2011} and many other complex processes \cite{Liew2010, Serafini1996, Xing2000}.

Beyond computer algorithms, other examples of large workflows that can be partitioned into smaller workloads are the transmission of big files over the Internet \cite{Wischik2011}, the processing of very large printing jobs using more than one printer, the introduction of additional roads in urban traffic \cite{Li2015, Silva2015, Vardi1996} and the breakup of manufacturing processes into parallel streams \cite{Ghazvini1998}. In all these cases, once all the workloads are processed, the results are pieced together to produce a useful output.

The parallelization procedure entails a decision on how to partition the workflow so that its full completion process takes the shortest time with minimum uncertainty. Uncertainty is a relevant and important variable because of the unavoidable fluctuations in processing a workload that each processing unit, channel or virtual machine undergoes when having to time share with other processes. This introduces a stochastic component into the execution of any program, which at times can actually increase the time it takes for a given workflow to finish\footnote{This is unlike map-reduce, which splits the execution inputs into equal parts. While map-reduce can minimize execution times it does not necessarily minimize uncertainty. Thus, while on average, execution times are reduced, single instances can still take very long times to process.}. Thus the need to incorporate these fluctuations into the partitioning procedure, so that the overall workflow completes in shorter times than the original non-partitioned one. 

In what follows, we describe a novel procedure for breaking any complex workflow into several workloads in such a way that once their outputs are joined, their full completion takes less time and uncertainty than when running it in only one processor. The procedure is based on notions of risk from economics that are used to combine primitive algorithms into new programs that are preferable to any of the primitive ones \cite{Huberman1997, Xu2008}. In our case however, we focus both on speeding up the completion time and lowering the uncertainty of the \textit{joint} execution of the complementary workloads. This implies that the overall processing time of the full program is determined by the longest running process. After presenting the method, we demonstrate its effectiveness  in two different scenarios; the optimization of a convex function and the transmission of a large computer file over the Internet

In order to handle this problem, consider a workflow $D$, partitioned into two workloads $D_i$ and $D_j$, each of which computes on different machines $i$ and $j$, with different processing speeds and fluctuating performances. Once the slowest workload has completed, the two outputs are joined together and the workflow is considered complete.

For simplicity in the exposition, we will assume that the completion time, $t_i$ for the full workflow $D$ executing on machine $i$ is a continuous variable which is Normally distributed with mean $\mu_i$ and standard deviation $\sigma_i$,
\[ p(t_i | D, \mu_i, \sigma_i) \sim \mathcal{N}(\mu_i, \sigma_i^2) \]

If the workload on machine $i$, $D_i$, is smaller than $D$ by a factor of $f$, i.e. $|D_i| = f|D|$, the resulting distribution of completion times $t_i$ for machine $i$ is given by,
\[ p(t_i | D_i, \mu_i, \sigma_i) \sim \mathcal{N} \left( f \mu_i, [f \sigma_i]^2 \right) \]
and similarly for machine $j$ that processes workload $D_j$, so that $|D_j| = (1 - f) |D|$, 
\[ p(t_j | D_j, \mu_j, \sigma_j) \sim \mathcal{N} \left( [1-f] \mu_j, \left[ (1-f) \sigma_j \right]^2 \right) \]

The workflow only completes when both machines $i$ and $j$ finish processing their assigned workloads $D_i, D_j$, respectively. Thus, the cumulative density function for the completion time $t$ is the probability that both workloads $t_i$ and $t_j$ complete within a time $\epsilon$.
\begin{align}
	\label{eqn:parallel_t}
	P(t \leq \epsilon | f, D, \mu_i, \sigma_i, \mu_j, \sigma_j) &= P(t_i \leq \epsilon | D_i, \mu_i, \sigma_i) \cdot P(t_j \leq \epsilon | D_j, \mu_j, \sigma_j)
\end{align}

The decision as how to partition the workflow consists in choosing the value of $f$ such that the workflow will execute with the lowest expected completion time $\mu(f)$ and variance $\sigma^2(f)$. This requires understanding the behavior of $\mu$ and $\sigma^2$ as a function of $f$. They can be derived from their probability density function as,
\begin{gather*}
	\mu(f) = E(t | \Theta) = \int_{0}^{\infty} t \cdot p(t | \Theta) ~ dt \\
	\sigma^2(f) = Var(t | \Theta) = \int_{0}^{\infty} t^2 \cdot p(t | \Theta) ~ dt - \left[ E(t | \Theta) \right]^2
	\end{gather*}
with
\[ \Theta \equiv \{ f, D, \mu_i, \sigma_i, \mu_j, \sigma_j \} \]
with the probability density function given by the first order derivative of the cumulative density function shown in Equation \ref{eqn:parallel_t},
\[ p(t | \Theta) = \frac{d}{dt} P(t' \leq t | \Theta) \]
Since there is no closed form solution for the probability density function, we express the expected completion time $\mu(f)$ in terms of the cumulative density function shown in Equation \ref{eqn:parallel_t},
\[ \mu(f) = \int_0^{\infty} 1 - P(t \leq \epsilon | \Theta) ~ d\epsilon \]
Similarly, the variance $\sigma^2(f)$ is given by,
\[ \sigma^2(f) = \left\{ 2 \int_0^{\infty} \epsilon \Big[ 1 - P(t \leq \epsilon | \Theta) \Big] ~ d\epsilon \right\} - \Big[ \mu(f) \Big]^2 \]

\begin{figure}[htb]
	\centering
	\subfloat[$\mu$ with respect to $f$]
	{
		\includegraphics[width=2.3in]{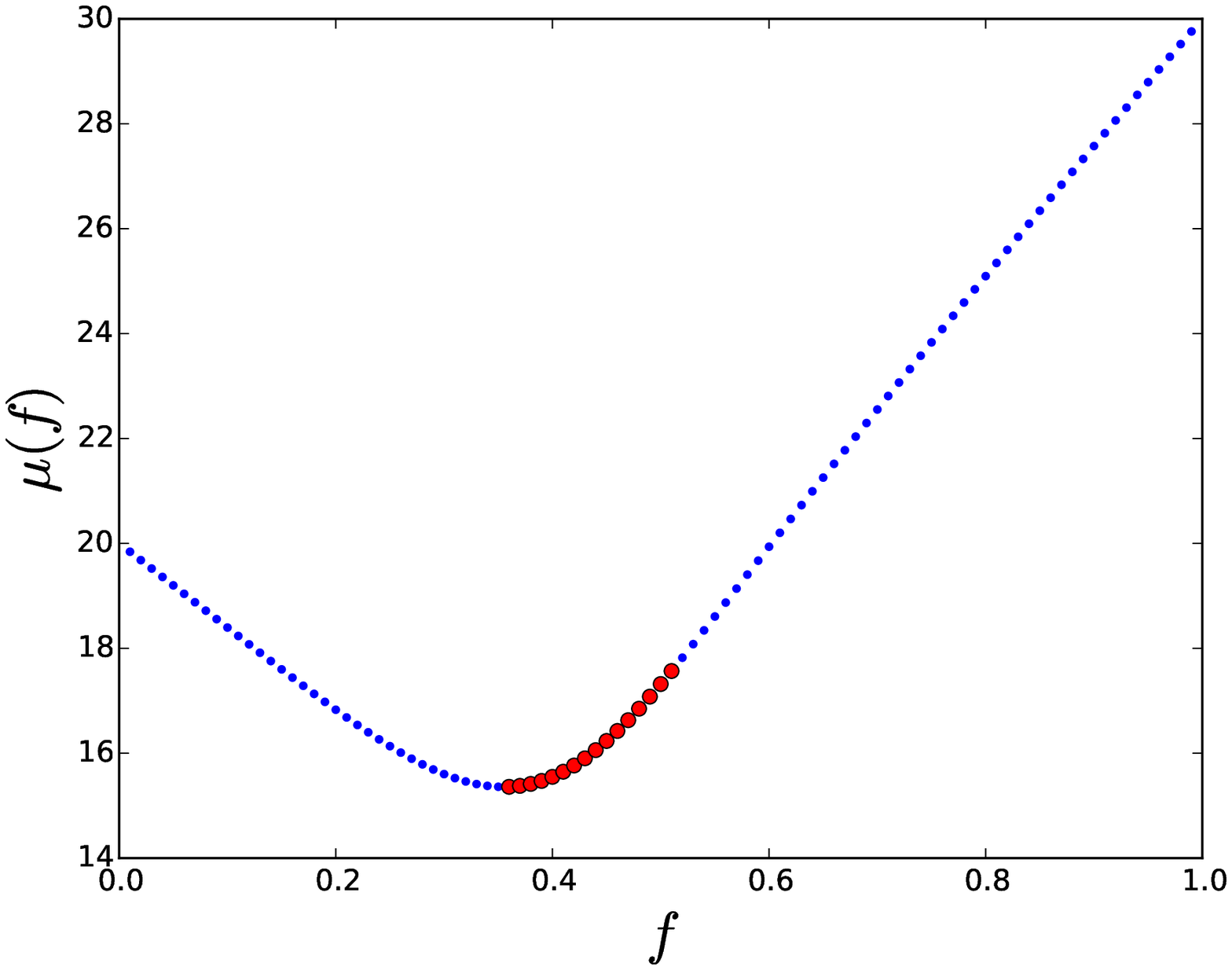}
		\label{fig:mu_vs_f}
	}
	\subfloat[$\sigma^2$ with respect to $f$]
	{
		\includegraphics[width=2.3in]{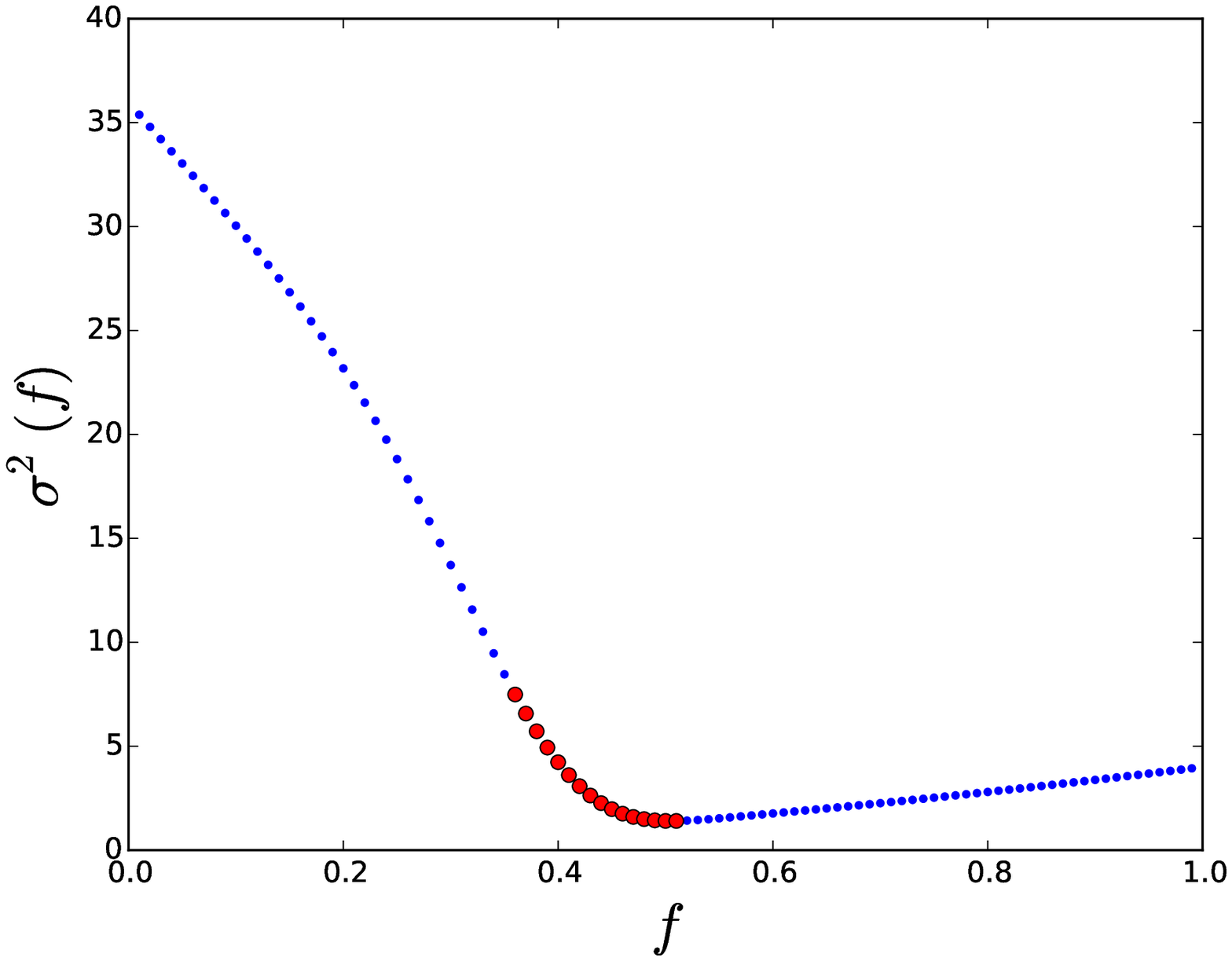}
		\label{fig:sigma2_vs_f}
	}
	\caption{Figures \ref{fig:mu_vs_f} and \ref{fig:sigma2_vs_f} show $\mu$ and $\sigma^2$ as a function of $f$. The values used are $\mu_i = 30$, $\sigma_i = 2$, $\mu_j = 20$, $\sigma_j = 6$. The bolded red $\bullet$ is the efficient frontier that provides the best combinations of $\mu$ and $\sigma^2$. The value of $f$ which gives the minimum point in each of (\ref{fig:mu_vs_f}) and (\ref{fig:sigma2_vs_f}) is different, and that results in a range of values that $f$ can possibly take.}
	\label{fig:portfolio}
\end{figure}

\begin{figure}[htb]
	\centering
	\includegraphics[width=4.0in]{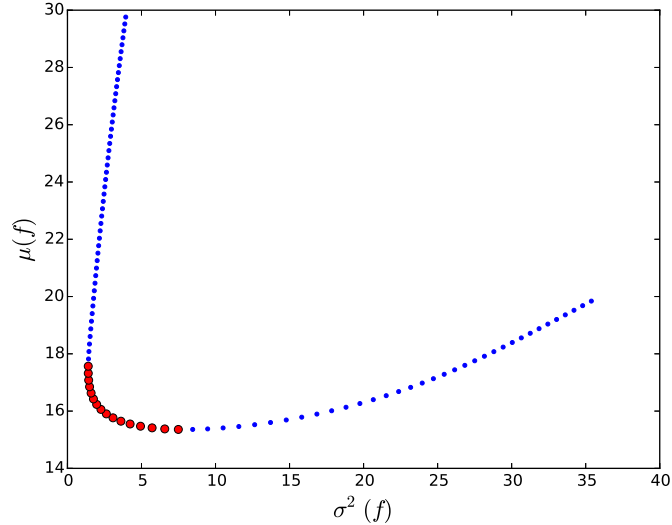}
	\caption{Parametric plot of $\mu$ and $\sigma^2$ for $\mu_i = 30$, $\sigma_i = 2$, $\mu_j = 20$, $\sigma_j = 6$. The bolded red $\bullet$ corresponds to the efficient frontier in Figures \ref{fig:mu_vs_f} and \ref{fig:sigma2_vs_f}.}
	\label{fig:mu_vs_sigma2}
\end{figure}

Figures \ref{fig:mu_vs_f} and \ref{fig:sigma2_vs_f} show the behavior of the expected time $\mu$ to completion of a workflow and its variance $\sigma^2$ both as a function of $f$; while Figure \ref{fig:mu_vs_sigma2} shows $\mu$ and $\sigma$ parametrically as a function of each other. As can be seen, the partition of the workflow results in completion times and variances that can be much smaller than the original unpartitioned ones. Moreover, the minima for $\mu$ and $\sigma^2$ occur for different values of $f$, a fact that determines a range of choices.

Since the resulting curve in Figure \ref{fig:mu_vs_sigma2} is parabolic, some values of $\mu$ have two possible choices of $\sigma^2$ and vice-versa. If our assumptions on the statistical distribution of completion times for the two parallel workloads hold, the theoretical results derived in Figure \ref{fig:mu_vs_sigma2} allows us to decide the appropriate value of $f$ which minimizes $\mu$ and $\sigma^2$ for the full workflow execution.

This methodology is general enough so as to be applicable to a number of scenarios. In what follows we illustrate this approach with two concrete examples that can be easily tested in the laboratory. The first one is the mathematical optimization of a convex function, while the second corresponds to the transmission of a large file over the Internet.

We first demonstrate the parallel optimization of a least squares error function used for logistic regression classification. This function is quadratic and therefore convex. This is different from a parallel algorithm such as map-reduce, which breaks the file into an equal number of smaller inputs. In our case the input data $D$ to the convex function is partitioned into two workloads of unequal sizes $D_i$ and $D_j$. A classical optimization algorithm \cite{Battiti1992} is then applied to each of the workloads $D_i$ and $D_j$ to obtain globally optimal solutions $\theta_i$ and $\theta_j$, based on each of their inputs. The desired solution $\theta$ for the original workflow $D$ was then obtained as a linear combination of the solution from each of the workloads.
\[ \theta = f \theta_i + (1-f) \theta_j \]

\begin{figure}[htb]
	\centering
	\subfloat[$\mu$ with respect to $f$]
	{
		\includegraphics[width=2.3in]{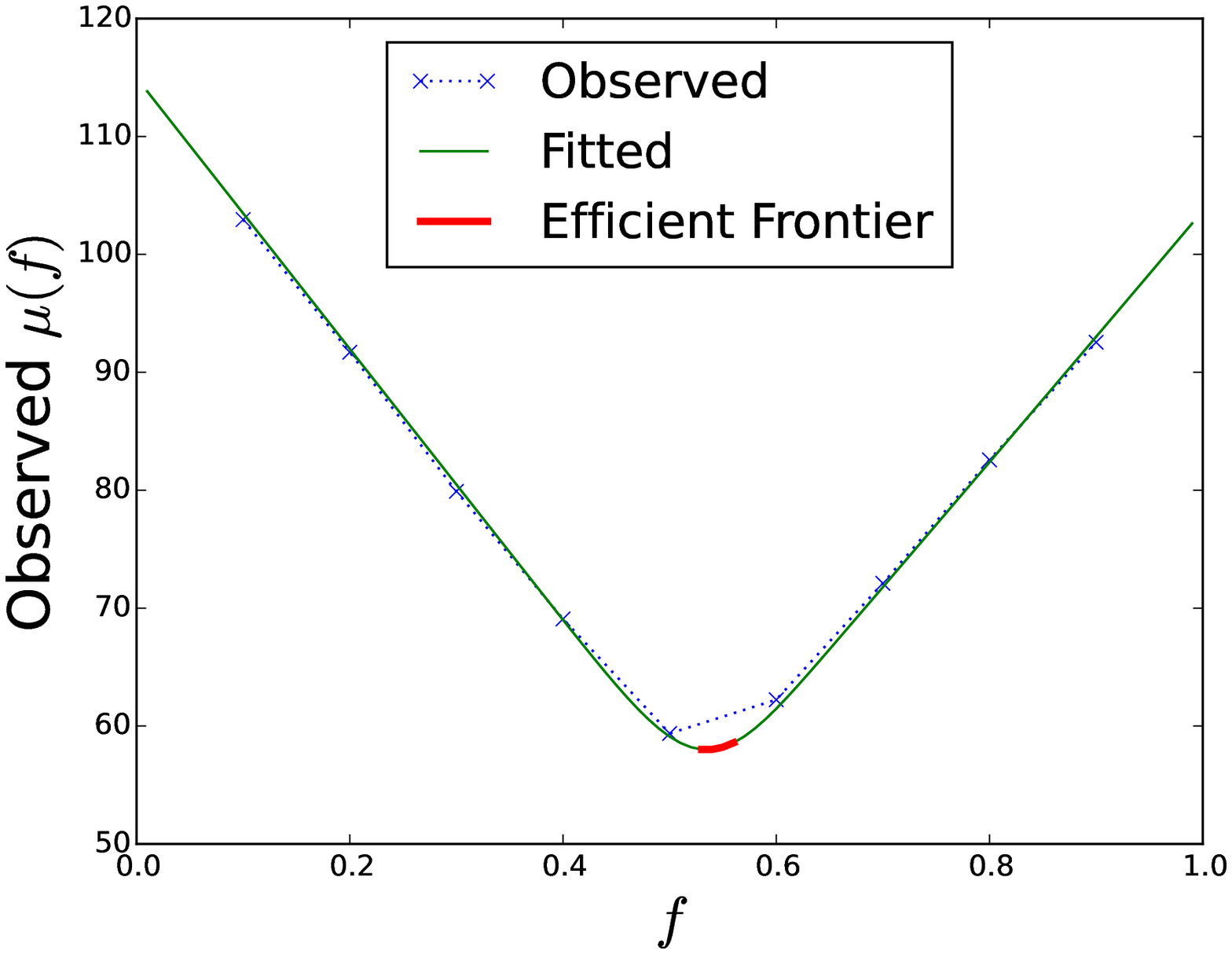}
		\label{fig:psgd15_observed_mu_vs_f_1}
	}
	\subfloat[$\sigma^2$ with respect to $f$]
	{
		\includegraphics[width=2.3in]{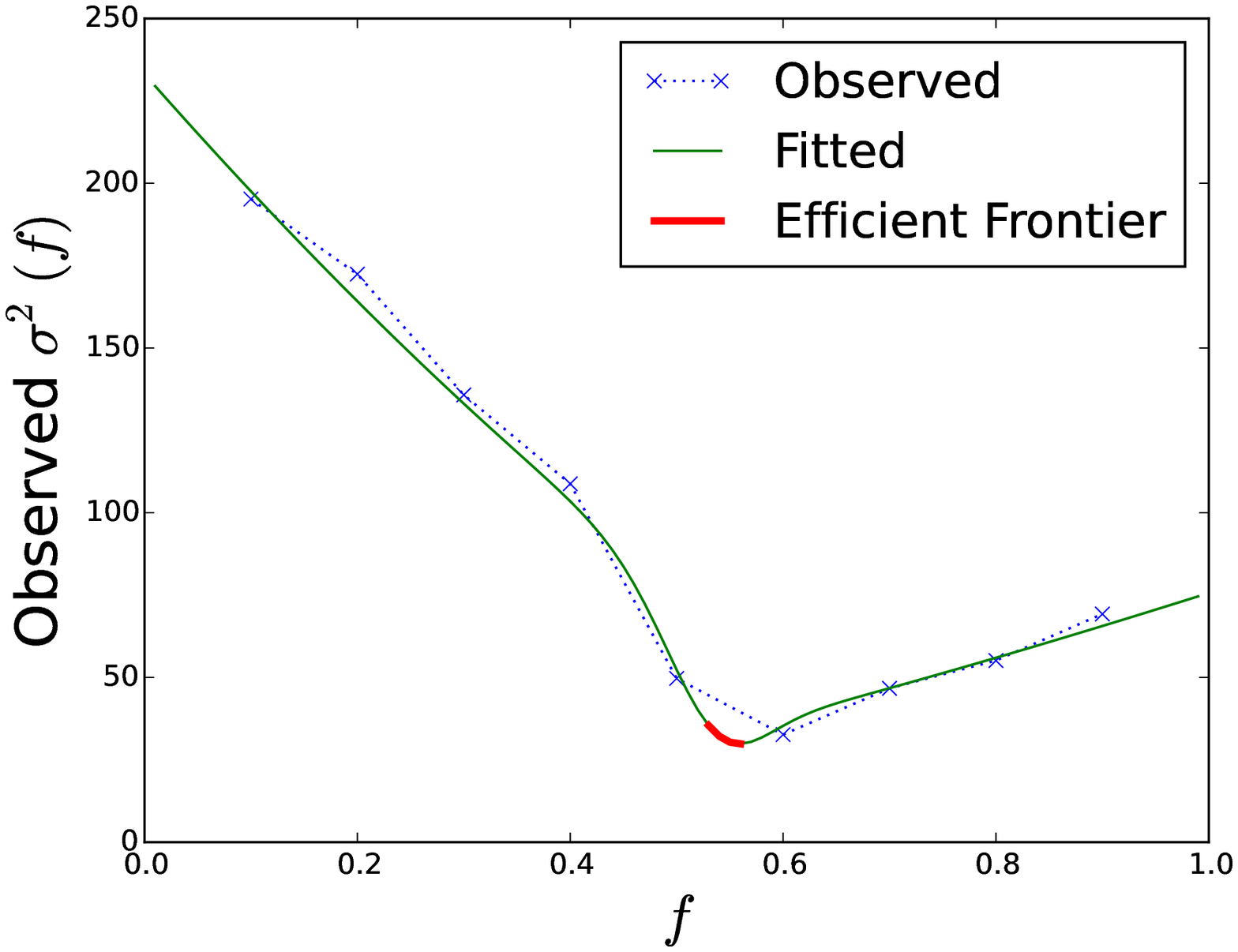}
		\label{fig:psgd15_observed_sigma2_vs_f_1}
	}
	\caption{$\mu$ and $\sigma^2$ as a function of $f$ for parallel optimization of the convex least squares error function.}
	\label{fig:psgd15_observed_mu_vs_f}
\end{figure}

\begin{figure}[htb]
	\centering
	\includegraphics[width=4.0in]{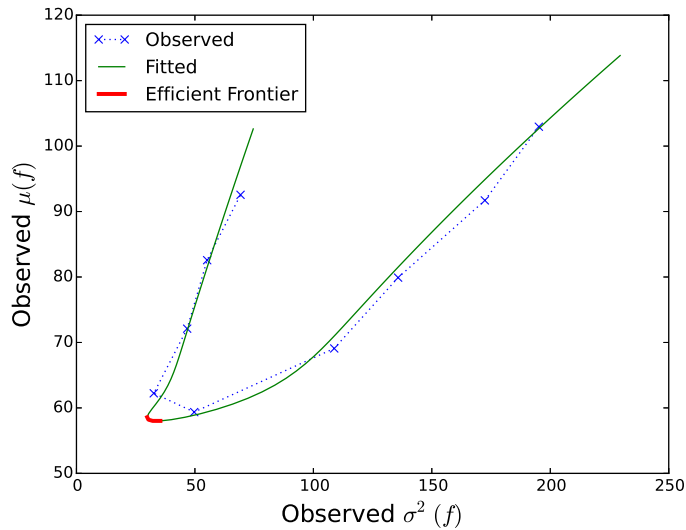}
	\caption{Observed values of $\mu(f)$ and $\sigma^2(f)$ as a parametric function of each other.}
	\label{fig:psgd15_observed_mu_vs_sigma2}
\end{figure}

We processed the parallel optimization algorithm on two virtual machines, each with one CPU core running at 2667 MHz. To generate uncertainty in the completion time of each workload, we ran background processes on each machine, which created contention for CPU resources. 

Figures \ref{fig:psgd15_observed_mu_vs_f_1} and \ref{fig:psgd15_observed_sigma2_vs_f_1} show how the mean completion times and their variances vary with each value of $f$. The mean and variance at each value of $f$ was obtained by repeating many trials of the optimization process over a long period of time using different values of $f$. Figure \ref{fig:psgd15_observed_mu_vs_sigma2} shows $\mu$ and $\sigma^2$ parametrically as a function of each other for this parallel optimization case. As can be seen, one obtains a performance curve similar to the theoretical one in Figure \ref{fig:mu_vs_sigma2}. More importantly, the results clearly show that both the total completion time and its variability are much lower than the original unpartitioned workflow. This implies that one can always choose a partition (given by the value of $f$) such that it lowers both the completion time of the computation and its uncertainty.

Next, we performed a file transmission experiment by transferring a fixed size file in parallel from a source node to a destination node over two network paths. Besides its intrinsic value, this experiment also acts as a proxy for other spatial workflows which are harder to test in the laboratory, such as urban traffic or transportation routes.  For our file transmission experiment, since the TCP network protocol does not allow fine grain control of how the file packets travel through the Internet, we created an intermediate overlay to redirect a fraction of the file packets through a different path.

The source node used in our experiment was hosted in New York City, while the destination node was hosted in Singapore. The use of traceroute showed that network packets went through the west coast of the US before reaching the destination in Singapore. This implies that network packets from New York City route through the Pacific Ocean to Singapore. 

We wanted to determine if having an alternate route for sending some of the file packets from New York to Singapore via Europe provided a better transmission process. We thus created another host in London to act as an overlay which received file packets from New York and forwarded them to Singapore. 

We split a large file into two workloads whose sizes depended on the prdifferent values of $f$. and sent each of them across the two different network channels. We ensured that only network transmission times contributed to the completion times by ignoring disk I/O delays.

\begin{figure}[htb]
	\centering
	\includegraphics[width=4.0in]{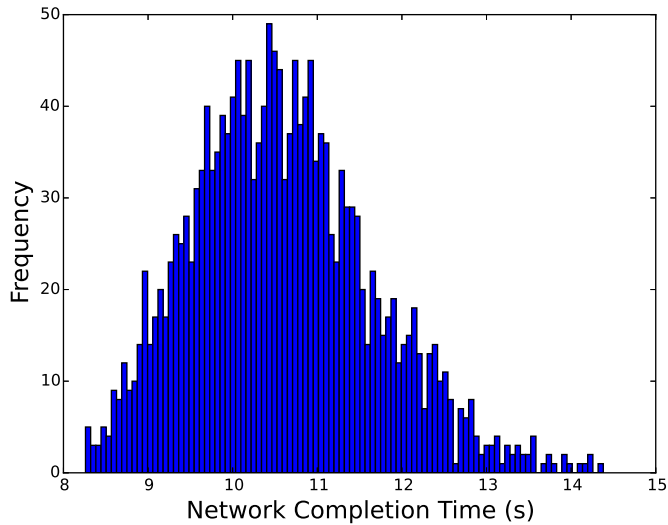}
	\caption{Histogram of completion times at $f=0.5$ for the one workload. Due to inherent fluctuations in the network pipelines, the completion time for a file of fixed size was Normally distributed around a mean and variance. This distribution of completion times was consistent for the values of $f \in \{0.0,0.1,0.2,\ldots,1.0\}$ during our experiments.}
	\label{fig:digitalocean_times_histogram}
\end{figure}

\begin{figure}[htb]
	\centering
	\subfloat[$\mu$ with respect to $f$]
	{
		\includegraphics[width=2.3in]{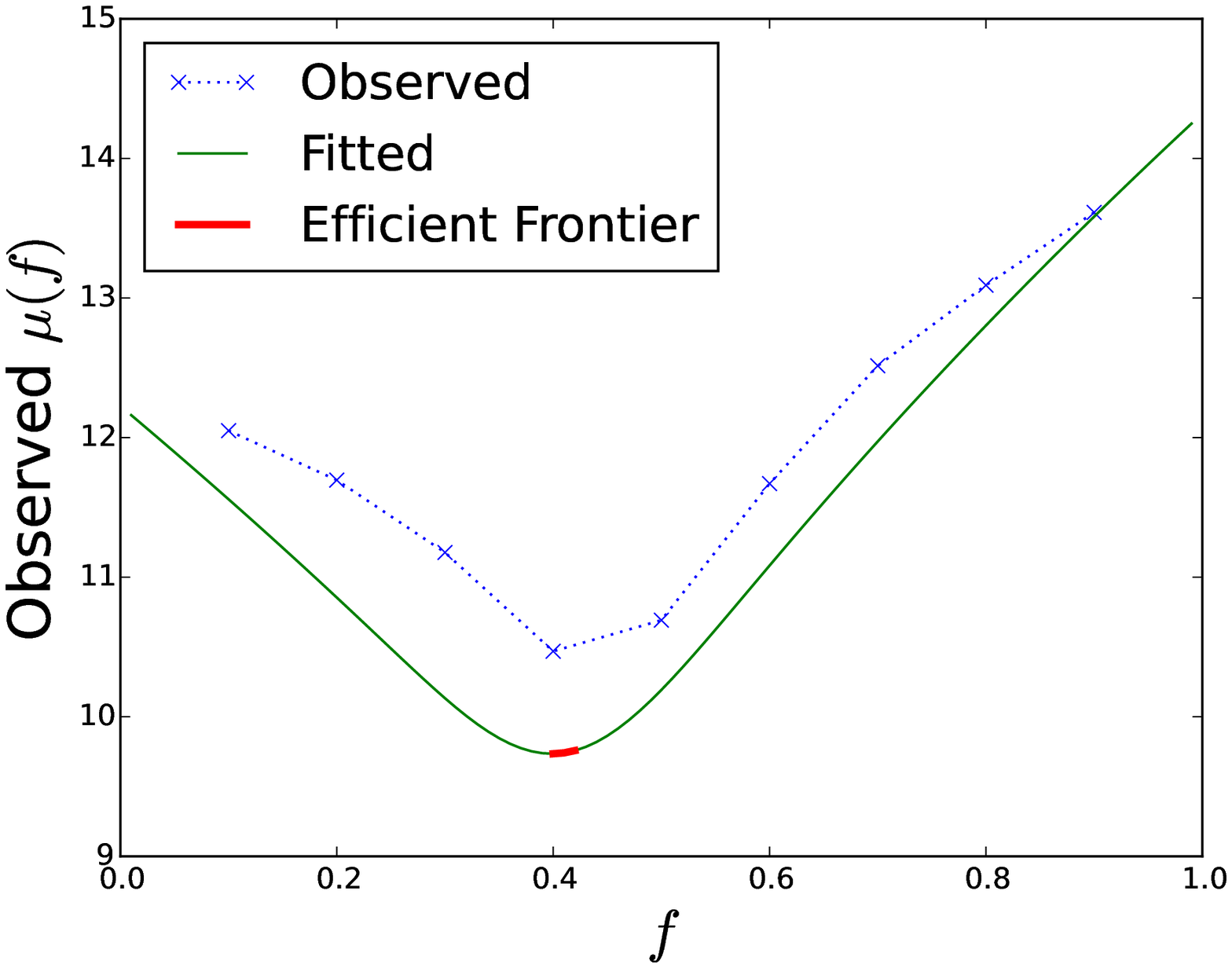}
		\label{fig:digitalocean_observed_mu_vs_f_1}
	}
	\subfloat[$\sigma^2$ with respect to $f$]
	{
		\includegraphics[width=2.3in]{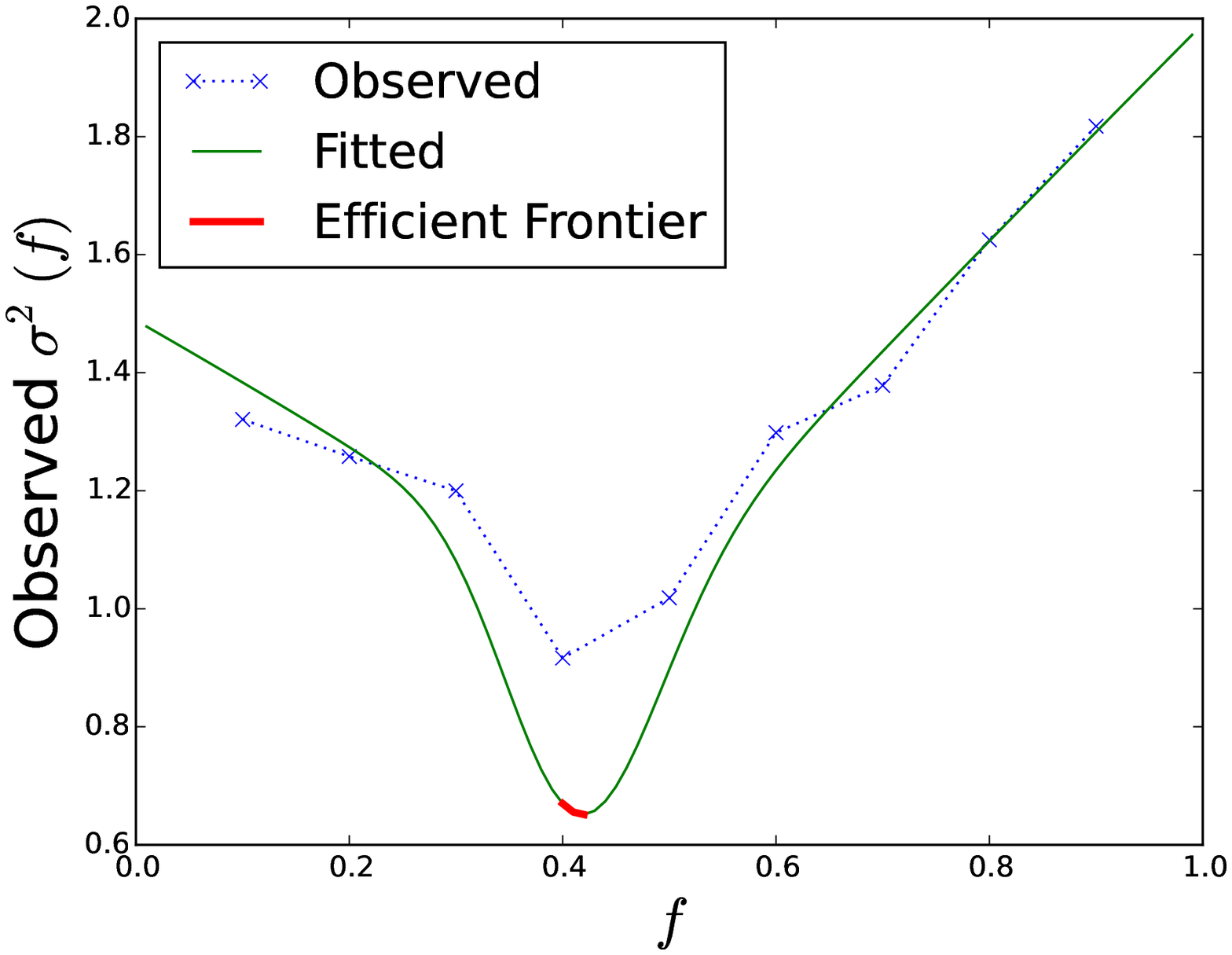}
		\label{fig:digitalocean_observed_sigma2_vs_f_1}
	}
	\caption{$\mu$ and $\sigma^2$ as a function of $f$ for dual transmission of a file.}
	\label{fig:digitalocean_observed_mu_vs_f}
\end{figure}
To measure the completion time of the two parallel file transfers, the node at the destination measured the time of the last packet (from either channel) and then subtracted the time of the request for the first packet (from both channels).

In order to measure the mean and variance of the transmission times, we repeated the file transfer 20024 times over a period of 72 hours from Sunday to Tuesday. For each trial, we randomized the value of $f$.

Figure \ref{fig:digitalocean_times_histogram} shows the distribution of completion times for the value $f=0.5$, which is well approximated by a Normal distribution. Figures \ref{fig:digitalocean_observed_mu_vs_f_1} and \ref{fig:digitalocean_observed_sigma2_vs_f_1} show how the mean completion times and their variances varied as a function of $f$. Similar to the optimization case, the results for this file transmission experiment are also consistent with the theoretical predictions shown in in Figures \ref{fig:mu_vs_f} and \ref{fig:sigma2_vs_f}.

These results show that this general methodology for partitioning uncertain workflows leads to shorter expected completion times with reduced uncertainty. All is needed after obtaining such a curve is to decide on the value of $f$ that lowers uncertainty and expected completion time. A very direct application of this method would be in the information technology domain, as it allows for new formulations of pricing schemes for Quality-of-Service (QoS) \cite{DaSilva2000, Ward2015} offerings, since in order to satisfy demand large cloud and data systems need to increase the speed with which they process incoming jobs.

There are several obvious extensions of this work. Unlike the scenarios we studied, where the statistical properties of the system are known, one often encounters situations where the processing capabilities of the systems have to be estimated on-the-fly. Methods based on Bayesian inference during deployment \cite{Murphy2007} would then provide the distribution in completion times that are needed to partition a given workload. Moreover, one can generalize the splitting procedure to very many components.In that case, methods like group testing \cite{Dorfman1943, Mezard2011} could be utilized to decide on the best choice of the number of components. 

Finally, we stress that the applicability of this method extends beyond the execution of computer algorithms and file transmissions over the Internet. Alleviating congestion in urban traffic, job scheduling in manufacturing, finding optimal routes for supply chain scenarios and any other activities that allow for some parallelism can also exploit this approach.

\newpage

\bibliographystyle{unsrt}
\bibliography{splitting}

\begin{thebibliography}{10}

\bibitem{Dean2008}
Jeffrey Dean and Sanjay Ghemawat.
\newblock Mapreduce: Simplified data processing on large clusters.
\newblock {\em Commun. ACM}, 51(1):107--113, January 2008.

\bibitem{Chu2007}
Cheng tao Chu, Sang~K. Kim, Yi~an~Lin, Yuanyuan Yu, Gary Bradski, Kunle
  Olukotun, and Andrew~Y. Ng.
\newblock Map-reduce for machine learning on multicore.
\newblock In B.~Sch\"{o}lkopf, J.C. Platt, and T.~Hoffman, editors, {\em
  Advances in Neural Information Processing Systems 19}, pages 281--288. MIT
  Press, 2007.

\bibitem{Low2012}
Yucheng Low, Danny Bickson, Joseph Gonzalez, Carlos Guestrin, Aapo Kyrola, and
  Joseph~M. Hellerstein.
\newblock Distributed graphlab: A framework for machine learning and data
  mining in the cloud.
\newblock {\em Proc. VLDB Endow.}, 5(8):716--727, April 2012.

\bibitem{Xing2015}
Eric~P. Xing, Qirong Ho, Wei Dai, Jin~Kyu Kim, Jinliang Wei, Seunghak Lee, Xun
  Zheng, Pengtao Xie, Abhimanu Kumar, and Yaoliang Yu.
\newblock Petuum: A new platform for distributed machine learning on big data.
\newblock In {\em Proceedings of the 21th ACM SIGKDD International Conference
  on Knowledge Discovery and Data Mining}, KDD '15, New York, NY, USA, 2015.
  ACM.

\bibitem{Zinkevich2010}
Martin Zinkevich, Markus Weimer, Lihong Li, and Alex~J. Smola.
\newblock Parallelized stochastic gradient descent.
\newblock In J.D. Lafferty, C.K.I. Williams, J.~Shawe-Taylor, R.S. Zemel, and
  A.~Culotta, editors, {\em Advances in Neural Information Processing Systems
  23}, pages 2595--2603. Curran Associates, Inc., 2010.

\bibitem{Brunner2009}
M.~Brunner, D.~Dudkowski, C.~Mingardi, and G.~Nunzi.
\newblock Probabilistic decentralized network management.
\newblock In {\em Integrated Network Management, 2009. IM '09. IFIP/IEEE
  International Symposium on}, pages 25--32, June 2009.

\bibitem{John2013}
W.~John, K.~Pentikousis, G.~Agapiou, E.~Jacob, M.~Kind, A.~Manzalini, F.~Risso,
  D.~Staessens, R.~Steinert, and C.~Meirosu.
\newblock Research directions in network service chaining.
\newblock In {\em Future Networks and Services (SDN4FNS), 2013 IEEE SDN for},
  pages 1--7, Nov 2013.

\bibitem{Prieto2011}
A.G. Prieto, D.~Gillblad, R.~Steinert, and A.~Miron.
\newblock Toward decentralized probabilistic management.
\newblock {\em Communications Magazine, IEEE}, 49(7):80--86, July 2011.

\bibitem{Liew2010}
Chee~Sun Liew, Malcolm~P. Atkinson, Jano~I. van Hemert, and Liangxiu Han.
\newblock Towards optimising distributed data streaming graphs using parallel
  streams.
\newblock In {\em Proceedings of the 19th ACM International Symposium on High
  Performance Distributed Computing}, HPDC '10, pages 725--736, New York, NY,
  USA, 2010. ACM.

\bibitem{Serafini1996}
Paolo Serafini.
\newblock Scheduling jobs on several machines with the job splitting property.
\newblock {\em Operations Research}, 44(4):617--628, 1996.

\bibitem{Xing2000}
Wenxun Xing and Jiawei Zhang.
\newblock Parallel machine scheduling with splitting jobs.
\newblock {\em Discrete Applied Mathematics}, 103(1–3):259 -- 269, 2000.

\bibitem{Wischik2011}
Damon Wischik, Costin Raiciu, Adam Greenhalgh, and Mark Handley.
\newblock Design, implementation and evaluation of congestion control for
  multipath tcp.
\newblock In {\em Proceedings of the 8th USENIX Conference on Networked Systems
  Design and Implementation}, NSDI'11, pages 99--112, Berkeley, CA, USA, 2011.
  USENIX Association.

\bibitem{Li2015}
Daqing Li, Bowen Fu, Yunpeng Wang, Guangquan Lu, Yehiel Berezin, H.~Eugene
  Stanley, and Shlomo Havlin.
\newblock Percolation transition in dynamical traffic network with evolving
  critical bottlenecks.
\newblock {\em Proceedings of the National Academy of Sciences},
  112(3):669--672, 2015.

\bibitem{Silva2015}
Ricardo Silva, Soong~Moon Kang, and Edoardo~M. Airoldi.
\newblock Predicting traffic volumes and estimating the effects of shocks in
  massive transportation systems.
\newblock {\em Proceedings of the National Academy of Sciences},
  112(18):5643--5648, 2015.

\bibitem{Vardi1996}
Y.~Vardi.
\newblock Network tomography: Estimating source-destination traffic intensities
  from link data.
\newblock {\em Journal of the American Statistical Association},
  91(433):365--377, 1996.

\bibitem{Ghazvini1998}
Fariborz~Jolai Ghazvini and Lionel Dupont.
\newblock Minimizing mean flow times criteria on a single batch processing
  machine with non-identical jobs sizes.
\newblock {\em International Journal of Production Economics}, 55(3):273 --
  280, 1998.

\bibitem{Huberman1997}
B.~A. Huberman, R.~M. Lukose, and T.~Hogg.
\newblock An economics approach to hard computational problems.
\newblock {\em Science}, 27:51--53, 1997.

\bibitem{Xu2008}
Lin Xu, Frank Hutter, Holger~H. Hoos, and Kevin Leyton-Brown.
\newblock Satzilla: Portfolio-based algorithm selection for sat.
\newblock {\em J. Artif. Int. Res.}, 32(1):565--606, June 2008.

\bibitem{Battiti1992}
R~Battiti.
\newblock First- and second-order methods for learning: Between steepest
  descent and newton's method.
\newblock {\em Neural Computation}, 4(2):141--166, March 1992.

\bibitem{DaSilva2000}
Luiz~A. DaSilva.
\newblock Pricing for qos-enabled networks: A survey.
\newblock {\em Communications Surveys Tutorials, IEEE}, 3(2):2--8, Second 2000.

\bibitem{Ward2015}
Julie Ward, Filippo Balestrieri, and Bernardo~A. Huberman.
\newblock Revenue management in cloud computing.
\newblock In {\em INFORMS Revenue Management and Pricing Conference}, June
  2015.

\bibitem{Murphy2007}
Kevin~P. Murphy.
\newblock Conjugate bayesian analysis of the gaussian distribution.
\newblock Technical report, University of British Columbia, 2007.

\bibitem{Dorfman1943}
Robert Dorfman.
\newblock The detection of defective members of large populations.
\newblock {\em The Annals of Mathematical Statistics}, 14(4):pp. 436--440,
  1943.

\bibitem{Mezard2011}
M.~M\'{e}zard and C.~Toninelli.
\newblock Group testing with random pools: Optimal two-stage algorithms.
\newblock {\em Information Theory, IEEE Transactions on}, 57(3):1736--1745,
  March 2011.

\end{thebibliography}

\end{document}